\documentclass[prb,preprint]{revtex4}
\usepackage{amsmath}  
\usepackage{amsfonts} 
\usepackage{graphicx} 
\usepackage{hyperref}

\newcommand{\bal}{\begin{align}}
\newcommand{\eal}{\end{align}}
\newcommand{\beq}{\begin{eqnarray}}
\newcommand{\eeq}{\end{eqnarray}}
\newcommand{\nneeq}{\nonumber \end{eqnarray}}

\newcommand{\nn}{\nonumber \\}
\newcommand{\es}{& = &}
\newcommand{\rs}{\, = \,}

\begin{document}

\title{ Elementary example of energy and momentum of 
        an extended physical system in special relativity }
\author{ Kamil Serafin }
\affiliation{ Institute of Theoretical Physics,
              Faculty of Physics, University of Warsaw,
              Pasteura 5, 02-093 Warsaw, Poland }
\author{ Stanis{\l}aw D. G{\l}azek }
\affiliation{ Institute of Theoretical Physics,
              Faculty of Physics, University of Warsaw,
              Pasteura 5, 02-093 Warsaw, Poland, 
              and  \\ Department of Physics, Yale University,
              New Haven, Connecticut 06520, USA
              }
\date{ 21 April 2017 }

\begin{abstract}
An instructive paradox concerning classical description of 
energy and momentum of extended physical systems in special 
relativity theory is explained using an elementary example 
of two point-like massive bodies rotating on a circle in 
their center-of-mass frame of reference, connected by an 
arbitrarily light and infinitesimally thin string. Namely, 
from the point of view of the inertial observers who move 
with respect to the rotating system, the sums of energies 
and momenta of the two bodies oscillate, instead of being 
constant in time. This result is understood in terms of the 
mechanism that binds the bodies: the string contributes to 
the system total energy and momentum no matter how light it 
is. Its contribution eliminates the unphysical oscillations 
from the system total four-momentum. Generality of the
relativistic approach, applied here to the rotor example,
suggests that in every extended physical system its binding
mechanism contributes to its total energy and momentum.
\end{abstract}

\maketitle

\section{ Introduction }

Relativistic description of extended physical systems is more
complicated than the nonrelativistic one. Even if a system is
assumed known in its center-of-mass frame, description of it
in other frames is involved. The reason is that in special
relativity one cannot easily separate equations of motion into
equations for the relative motion of the system parts and equations
for the motion of the system as a whole.\cite{instant-front}
Relativity also leads to paradoxes concerning the total energy 
and momentum of system constituents. We discuss an elementary
example of such paradox. Our example is meant to be useful to
students of special relativity who try to apply the theory in
description of extended objects in motion. The example is also 
relevant to quantum description of bound states, but we do not 
discuss quantum theory.

The system we consider consists of two massive points connected
via an arbitrarily light and infinitesimally thin string, which 
may appear negligible as far as the total energy and momentum of 
the system are concerned. Held by the straight string, the massive 
points move around the center of mass of the system with constant 
angular velocity. We shall call this system a {\it rotor}, and 
often refer to the masses at the string ends as {\it endpoint masses},
located at the string {\it endpoints}. 

The issue of proper relativistic description of a rotor is not 
purely academic. Rotors are of interest in particle physics as 
models of mesons made of a quark, an anti-quark, and a string 
of gluons that connects them.\cite{Nambu, Goto, Susskind, ChodosThorn, 
GGRT, QCDString} The string prototype in a rotor is also useful 
in developing string theory as a candidate for explaining the 
nature of particles such as quarks themselves, in the quantum 
version of the string theory.\cite{Zwiebach,superstring} The 
need for understanding the rotor is illustrated by the fact 
that one can imagine that a string between quarks is always
straight,\cite{QCDString} while a relativistically acceptable 
picture of the rotor shows that the string bends when the 
rotor is in motion. Another field where a relativistic 
description of a moving rotor may be relevant is the 
astrophysics of binary systems, especially binary
pulsars,\cite{pulsars} although the relativistic effects 
we talk about are very small in them. They are also likely to 
be small in the collision of black holes that generates 
gravitational waves.\cite{gravitationalwaves}

The rotor has a feature which simplifies its analysis as
a relativistic extended system: there is a clear distinction
between its endpoint constituents, {\it i.e.}, the masses
at the string ends, and their binding mechanism -- the string.
In the Newtonian physics, if we assume that the string is very
light, then its contribution to the energy and momentum of
the system is small in every inertial frame of reference.
So, in the Newtonian physics, a very light and thin string
may appear negligible because it is responsible for no other
effect than the circular motion of the endpoint masses.

In the framework of special relativity the situation is 
different, because the laws of conservation of energy and 
momentum require a proper treatment of the string. The need 
for the proper treatment is visible in the case of a moving 
rotor. Namely, in a frame of reference, in which the rotor 
moves in the direction perpendicular to its axis of rotation, 
the total energy and momentum of the endpoint masses
oscillate in time. This effect will be explained below. We 
also show that the string contribution cancels the oscillation,
no matter how light the string is. The resulting constant 
rotor total four-momentum has the proper Lorentz transformation 
properties.

The reader may find it interesting and encouraging to think
about the rotor example, knowing that the Lorentz transformations 
of the energy and momentum of a moving system had been studied
before by Takagi.\cite{takagi} However, that study is in its
nature concerned only with average values of energy and momentum,
and the so-called ``minimal'' addition to energy-momentum of
constituents proposed by Takagi is a positive ``scalar type
volume energy,'' which does not contribute to the system
three-momentum in any frame of reference. Here, we assume
that the laws of conservation of energy and momentum are
obeyed in every instant of time in all inertial frames of
reference and we introduce a different type of addition -- one
which contributes to the energy as well as three-momentum.

The great feature of theory of relativity is that it leads
to unacceptable results unless the relationships between cause
and effect are properly included in description of motion of
an extended system. In the rotor case, it turns out that the
stresses in a string that connects the endpoint masses contribute
a negative amount to the total energy of the rotor in the frames
of reference that uniformly move with respect to the rotor.
Our reasoning is sufficiently general to say that a relativistic
binding mechanism may contribute not only to the energy but also
to the momentum of an extended system. Another way to point out
this feature of theory of relativity is to say that description
of {\it gedanken experiments}, the concept introduced by Einstein
to imagine what {\it in principle} may happen physically, leads
to errors unless the events considered in these experiments are
physically admissible in the sense that every effect has a cause.
Motion of a rotor can be imagined as a gedanken experiment and
teaches us about utility of classical relativity theory. For example,
one is forced to correct the description of endpoint masses by
including the string. Whatever the binding mechanism is, its
contribution to the energy and momentum of the system must be
included to avoid inconsistency.

It is known that relativistic description of extended systems
leads to paradoxes. The ladder paradox\cite{rindler1961} is
mentioned quite often. Less popular is Bell's spaceship
paradox,\cite{dewanberan} over which disputes continue
until today.\cite{bellsnote,bellscomment} The right-angle
lever paradox\cite{ralever,PanofskyPhillips} was discussed
by Laue.\cite{Laue} Pauli discussed the Trouton and Noble
experiment with a moving condenser.\cite{Pauli,TroutonNoble}

The rotor paradox discussed in this paper differs from the
classic paradoxes mentioned above. The latter concern extended
physical systems looked upon from the frames of reference that
uniformly move with respect to these systems, at least momentarily.
The rotor paradox concerns a physical system which cannot
ever be considered to be at rest in any inertial frame of
reference; it rotates.

We show on the rotor example that in such circumstances a
relativistic description of an extended system can be obtained
using the concept of energy-momentum tensor -- the space-time
density of system energy, momentum and stresses -- rather
than the concept of force. In relativistic description of 
extended systems that are more complex than the rotor of our
example, the energy-momentum tensor continues to be a useful
tool, worth study and practice. The rotor example is instructive
in this respect. 

Textbooks also mention the effect that the interaction
that binds constituents, such as the Coulomb interaction
that binds charges, contributes to the system energy, and
that the motion of the bound charges may lead to radiation.
For example, see Ref.~\onlinecite{PoissonWill}. In the
rotor, the binding is not necessarily electromagnetic.
Nevertheless, the rules of special relativity are powerful
enough to determine the form of energy-momentum tensor that
provides description of the binding mechanism in the rotor. 
The relativity dictum is general enough to imply that the
same treatment of energy-momentum and stresses applies to
all kinds of binding dynamics in extended systems.

For readers interested in further reading about physics of
extended systems we ought to mention that nonrelativistic
systems are described in standard textbooks of mechanics.
Classical string theory is instructively addressed in the
textbook by Zwiebach.\cite{Zwiebach} Generalization
of theory of one-dimensional string-like objects to three 
dimensional objects (relativistic elasticity theory) is 
provided by Kijowski and Magli.\cite{KijowskiMagli}
The string binding mechanism of an extended system that is
described here can be encoded in a Lagrangian density of
the form that generalizes the one given by Chodos and
Thorn.\cite{ChodosThorn} The generalization is required
when one allows for the string properties to vary along its
length, but the present article only briefly mentions
the Lagrangian approach for readers interested in the subject.

This article is organized in the following manner. In 
Sec.~\ref{sec:2masses}, we describe the paradox concerning 
the energy and momentum of the rotor when its description 
does not involve the string. Sec.~\ref{sec:string} is 
devoted to the construction of a complete energy-momentum
tensor of the rotor, including the string, and its analysis. 
We also present there examples of the rotor energy-momentum
tensors including stresses for different choices of the string
properties. Summary of our conclusions constitutes
Sec.~\ref{sec:conclusions}. In the Appendix we prove
an important theorem that vanishing four-divergence
of stress-energy tensor guarantees that four-momentum
is conserved and is a four-vector.

\section{Relativistic motion of two point-like masses}%
\label{sec:2masses}                                   %

One of the most important points of this article is
that the sum of four-momenta of the endpoint masses 
in a rotor cannot be identified in a relativistic 
theory with a correct energy-momentum of the rotor as 
a whole. This point is explained in this section.

In the center-of-mass frame of the rotor endpoint masses,
which we call the frame $R$, the masses by definition
move on a circle. Namely, the position of the first
mass is $r\vec n(t_R)$ and the position of the second
mass is $-r\vec n(t_R)$, where $r$ is the rotor radius
and $\vec n$ is a unit vector dependent on the time
$t_R$ in the frame $R$. We assume $\vec n$ in the form
\begin{equation}
\vec n(t_R) = \left[\begin{array}{c}
                     \cos\Omega t_R\\
                     \sin\Omega t_R\\
                     0
                    \end{array}
                    \right]\,,
\end{equation}
where $\Omega$ is the angular velocity of rotation.
In the frame $R$, the sums of energies and momenta of 
endpoint masses are conserved: the total energy equals
twice the constant energy of one mass moving on a circle,
and the total three-momentum is always zero. The observer
associated with the frame $R$ is called the observer $R$.

We now ask if the total four-momentum of endpoint masses
is also conserved from the point of view of observers
whose frames of reference move uniformly with respect
to the rotor. In order to check that, one  should transform
the positions of the endpoints to a moving frame, differentiate
them with respect to the time in the moving frame to get
the velocities and thus also energies and momenta, and
add respective energies and momenta of the masses at
a single instant of time in the moving frame. We describe
the resulting total energy-momentum of the endpoint masses
in the moving frame and we explain why it is not constant.

The inertial frame of reference of the observer moving with
respect to the rotor center of mass, is denoted by $M$.
The observer is called the observer $M$. We choose the
frame $M$ to move along $x$ axis of the frame $R$, with
velocity $-U$, so that the rotor as a whole moves with
velocity $+U$ along the $x$ axis of the frame $M$. Time
assigned to events in the moving frame, $t_M$, is connected
with the time assigned to the same events in the rest frame,
$t_R$, via the Lorentz transformation,
\begin{equation}
 c t_M = \gamma_U \left(c t_R + \beta_U x_R \right)\,,
 \label{lorentz-time}
\end{equation}
where $c$ is the speed of light in the vacuum, $\gamma_U =
(1-\beta_U^2)^{-1/2}$ is called the Lorentz gamma factor
and $\beta_U=U/c$.

\begin{figure}[h]
 \includegraphics{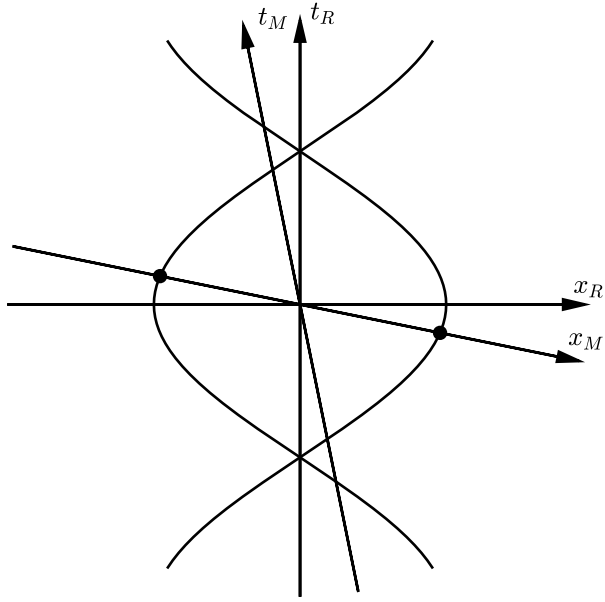}
 \caption{Curvy lines represent world lines of the endpoints,
          dots mark simultaneous measurement of energy-momenta
          of the masses in the moving frame (represented by
          $t_M$ and $x_M$ axes). In the rest frame (represented by
          $t_R$ and $x_R$ axes) the measurements are not simultaneous.
          Coordinates $t_M$, $x_M$ and $t_R$, $x_R$ are 
          related to each other via Lorentz transformation. 
         }
 \label{fig:times}
\end{figure}

Let us assume that the observer $M$, who is at rest at the
origin in the moving frame, measures energy-momenta of both
endpoint masses at the same time $t_M$. The observer $R$ at
rest in the rest frame, can see that $M's$ measurement on
the first mass was done at the moment $t_{1R}$, and on the
second mass at $t_{2R}$. Substituting $r\cos\Omega t_{1R}$,
which is the $x$ component of the position of the first mass,
for $x_R$ and $t_{1R}$ for $t_R$ in Eq.~(\ref{lorentz-time}),
one obtains $t_{1R}$ as a function of $t_M$. Similarly,
substituting $-r\cos\Omega t_{2R}$, which is the $x$ component
of the position of the second mass, for $x_R$ and $t_{2R}$
for $t_R$, one obtains $t_{2R}$ as a function of $t_M$.
We observe that if $\beta_U\neq0$, then $t_{1R}(t_M) \neq
t_{2R}(t_M)$, see Fig.~\ref{fig:times}.
Masses' energies and momenta at an instant of time in $M$
correspond to the masses' energies and momenta at different
instants of time in $R$, and hence their sum depends
on time in $M$.

This conclusion can also be arrived at in the following way.
The four-momenta of particles seen as separate can be transformed
individually from one inertial frame to another
using the Lorentz transformation (this is equivalent to saying
that they are four-vectors). Geometrically, these four-momenta
are assigned to points on the particles' world-lines. Their sum
depends on the choice of these points. Inertial observer's method
of choosing these points is to construct a hyperplane of simultaneity,
which is a set of the spacetime points whose coordinates have
the same time component, and finding the intersection points of
particles' world-lines with this hyperplane. Different inertial
observers choose different hyperplanes, according to the relativity
of simultaneity, and the total four-momentum of the endpoint masses
is time dependent in the frame $M$ despite that it is constant
in the frame $R$.

\begin{figure}[h]
 \includegraphics{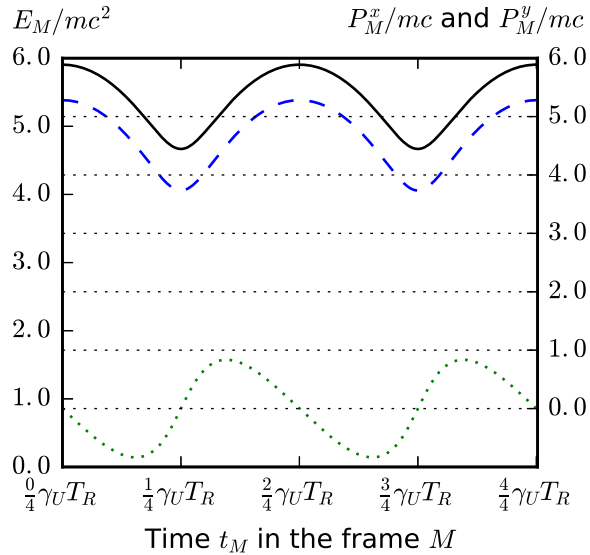}
  \caption{
   Oscillations in time $t_M$ of the total energy and
   total momentum of the rotor endpoint masses in the frame $M$.
   The total energy is represented by the solid line (black
   in the online version), marked on the left vertical axis,
   whereas $x$ and $y$ components of the total momentum are
   represented by dashed and dotted lines, respectively (blue
   and green, respectively in the online version), marked on
   the right vertical axis. Endpoints have masses $m=1$~kg each,
   the rotor radius is $r=1$~m, $v=-0.7c$ and $U=0.8c$.
   $T_R$ is the period of rotation in the frame $R$.
   See also Fig.~\ref{fig:momentum-hooke}.
  }
 \label{fig:oscillations}
\end{figure}

The result for energy and momentum of the endpoint masses
in the moving frame is,
\beq
E_M \es 2\gamma_U \gamma_v m c^2 + m U v \gamma_U \gamma_v
(s_2 - s_1) \ ,
\label{energy-moving}\\
\vec P_{M} \es 2 m \gamma_U \gamma_v
\begin{bmatrix}
U\\
0\\
0
\end{bmatrix}
+ m \gamma_v v
\begin{bmatrix}
\gamma_U (s_2 - s_1) \\
c_1 - c_2 \\
0
\end{bmatrix} \ ,
\label{momentum-moving}
\eeq
where $m$ is the mass of the rotor endpoint, $\Omega$
is the angular velocity of the rotation, and
\beq 
s_1 \es \sin[\Omega t_{1R}(t_M)] \ , \\
s_2 \es \sin[\Omega t_{2R}(t_M)] \ , \\
c_1 \es \cos[\Omega t_{1R}(t_M)] \ , \\
c_2 \es \cos[\Omega t_{2R}(t_M)] \ .
\eeq
Further, $\gamma_U$ and $\gamma_v$ are the Lorentz factors
corresponding to, respectively, the speeds $U$ and $v=\Omega r$.
The latter is the speed of each mass in the frame $R$.
Note that the first terms on the right-hand sides of
Eqs.~(\ref{energy-moving}) and (\ref{momentum-moving})
are precisely what one would obtain by Lorentz transforming
the total four-momentum of the masses from the frame $R$
to the frame $M$. The second terms in these equations
oscillate in time $t_M$. An example of these oscillations
in total energy and total momentum of the rotor endpoint
masses is presented in Fig.~\ref{fig:oscillations}.

\section{ The mechanism that binds the system }
\label{sec:string}

The oscillating energy of the rotor endpoint masses indicates
that our description of the system misses some ingredient,
which is needed in a theory meant to be relativistic. Indeed,
in Sec.~\ref{sec:2masses} we did not include in the description
any mechanism that binds the masses and forces them to move
on a circle. We can account for the missing ingredient by
describing the energy and momentum of the string that connects
the masses.

Since we analyze the energy and momentum of an extended body,
it is useful to employ energy-momentum tensor $T$. It has
sixteen components and depends on the point in space-time,
somewhat analogously to the relativistic electromagnetic
field tensor $F$ whose components are built from components
of the electric and magnetic fields, $\vec E$ and $\vec B$,
that may vary in time and space. $T$ is a symmetric tensor
of rank 2, whose $T^{00}$ component is energy density,
$T^{0i}$ component, $i = 1, 2, 3$, is the density of $i$-th
component of momentum multiplied by the speed of light
and $T^{ij}$ components form the stress tensor, {\it i.e.},
$T^{ij}$ is the density of flux of $j$-th component of momentum
through a surface perpendicular to $x^i$ axis. 

An introduction to phenomenology of stress-energy tensors 
can be found in Ref.~\onlinecite{Hartle}. Descriptions of 
canonical and symmetric stress-energy tensors are found 
in Refs.~\onlinecite{FieldTheoryLL, Soper}. A commonly known
example of a stress-energy tensor for extended systems
is the one for the perfect fluid,\cite{Hartle-fluid}
\beq
T^{\mu\nu}(x)
\es
\frac{1}{c^2} \epsilon u^\mu u^\nu
- p \left( \eta^{\mu\nu} - \frac{u^\mu u^\nu}{c^2} \right)
\ ,
\label{tmn-fluid}
\eeq
where $\epsilon$ is the local rest frame energy density
of the fluid, $p$ is the local rest frame pressure of the
fluid, $u^\mu$ is the four-velocity field of the fluid
and $\eta^{\mu\nu}=\text{diag}(1,-1,-1,-1)$ is the inverse of
the Minkowski metric matrix. $\epsilon$, $p$ and $u^\mu$
depend on the spacetime position $x$ and ``local rest frame''
of the fluid (assigned to spacetime point $x$) is an inertial
frame in which spatial components of $u^\mu(x)$ are zero.
In such frame, $T^{\mu\nu}=\text{diag}(\epsilon,p,p,p)$.
Note that $\eta^\mu_{\ \nu} - u^\mu u_\nu /c^2$ is the
projection operator on the subspace perpendicular (in
four-dimensional sense) to $u^\mu$. In applications, one
usually assumes some relation between $\epsilon$ and $p$
that is characteristic for the medium one wants to describe,
{\it e.g.}, a ``gas of photons'' in the cosmological models
is characterized by $\epsilon=3p$. An example of similar
relation for the string is given below in Eq.~(\ref{eps-p}).

The most important feature of stress-energy tensor
for our analysis is that, if the four-divergence
of the stress-energy tensor of a system vanishes,
\beq
\partial_\mu T^{\mu\nu} = 0\,,
\label{conservation-law}
\eeq
then the total four-momentum of the system,
\beq
\label{Pt}
P^\mu(t) = \frac{1}{c}\int d^3x\, T^{0\mu}(t,\vec x) \ ,
\eeq
is conserved. In other words, Eq.~(\ref{conservation-law})
is the locally defined condition that ensures the total
energy and momentum conservation laws: a divergenceless
tensor $T$ secures that $P$ is a four-vector that
does not depend on time. The proof is based on the
four-dimensional generalization of the Gauss's theorem:
integration of divergence of a field over a volume in which
it is enclosed is equal the flux of the field through the
surface of that volume. If the divergence vanishes in the
volume, the same amount of field flows in as flows out of
that volume. Since the extra fourth dimension is time,
and the flux through space describes the total four-momentum
of a system, the result that the same flux flows into
a space-time volume in the past as flows out of it in
the future, means that the four-momentum of the system
is conserved. For a complete prove we refer the reader
to the Appendix, where we also show  that, tensor nature
of $T$ guarantees that the total four-momentum is a Lorentz
four-vector, even though integration in Eq.~(\ref{Pt})
is over different spacetime hypersurfaces for different
inertial observers.

In Secs.~\ref{sec:string-stress-energy-tensor} and
\ref{sec:eqn-analysis} we first build the stress-energy
tensor of the string in frame R using physical intuition;
then we analyze it in frame M in Secs.~\ref{sec:en-mom-moving-frame},
\ref{sec:examples} and \ref{sec:ultrarelativistic}.

\subsection{ String energy-momentum tensor, including stresses }
\label{sec:string-stress-energy-tensor}

We construct the string energy-momentum tensor by using
an intuitive idea that the string is a chain of point-like
particles. Our first step is to consider the energy-momentum
tensor for one point-like particle. One tries first using
Eq.~(\ref{tmn-fluid}) with $\epsilon(t,\vec x\,)
= mc^2 \delta^{(3)}[\vec x- \vec r(t)]$ and $p=0$, but
this guess turns out to be wrong, giving wrong energy and
momentum of the particle when inserted into Eq.~(\ref{Pt}).
The problem is that the Dirac $\delta$-function needs to be
replaced by a properly transforming density. The correct
form of stress-energy tensor of one point-like particle is
\beq
T_{1}^{\mu\nu}(t_R,\vec x_R) 
\es
\frac{m}{\gamma_u}
\delta^{(3)}\left[\vec{x}_R - \vec r(t_R)\right]
u^\mu(t_R) u^\nu(t_R)\ ,
\label{t-1}
\eeq
where $m$ is the mass of the particle, $\vec r$ is 
it's position vector, $u^\mu$ it's four-velocity, 
and the Lorentz factor $\gamma_u$ corresponds to 
the speed 
\beq
u \es |\dot{\vec r}\, | \ .
\eeq
The Dirac $\delta$-function $\delta^{(3)}(\vec r = 
r_x \hat x + r_y \hat y + r_z \hat z)$, where $\hat x$,
$\hat y$ and $\hat z$ are three orthonormal basis 
vectors in space, is a product $\delta(r_x)\delta(r_y)\delta(r_z)$
and it locates the energy and momentum of the particle
at the point of it's position in space in the frame $R$
at the instant of time $t_R$. The factor $\gamma_u$ needs
to be included to describe the Lorentz contraction required
in a frame-dependent definition of the infinitesimal
spatial volume in which the point-particle is contained.
The $u^\mu u^\nu$ term is necessary so that one obtains
the proper expressions for the energy and momentum
of a particle by integrating $T_1$ in space. Namely,
integration of $T_{1}^{\mu 0}/c$ over $\vec x_R$ gives
$m u^\mu u^0/\gamma_uc = m u^\mu$, which is the correct
value of four-momentum of a point-like particle.
Furthermore, our expression for $T_{1}^{\mu\nu}$
is implicitly Lorentz covariant, which means that the
form of Eq.~(\ref{t-1}) remains the same in
any inertial reference frame with Cartesian spatial
coordinates.

Discussion of the stress-energy tensors for point-like
particles can be found in
Refs.~\onlinecite{LandauLifshitz-point-tmn,InfeldPlebanski}.
We use Eq.~(\ref{t-1}) to build the stress-energy tensor
of the rotor.

\subsubsection{ Part of tensor $T$ due to motion }

The first term is the stress-energy tensor of
the massive endpoints,
\begin{equation}
T_{m R}^{\mu\nu}(t_R,\vec x_R) =
\frac{m}{\gamma_{u_r}} \
\delta^{(3)}\left[\vec{x}_R - \vec r_1(t_R)\right]
u_1^\mu(t_R) u_1^\nu(t_R)
+
\frac{m}{\gamma_{u_r}} \
\delta^{(3)}\left[\vec{x}_R - \vec r_2(t_R)\right]
u_2^\mu(t_R) u_2^\nu(t_R)\,,
\label{t-m-r}
\end{equation}
where $\vec r_1(t_R) = r \vec n(t_R)$ and $\vec
r_2(t_R) = -r\vec n(t_R)$ are the positions of
the first and the second endpoint mass in the
frame $R$, respectively. The four-vectors
$u_1^\mu$ and $u_2^\mu$ are their respective
four-velocities.

The second part of the stress-energy tensor we construct
is the one, which should be present due to the motion
of parts of the string in the frame $R$. We introduce
\begin{equation}
T_{\epsilon R}^{\mu\nu}(t_R,\vec x_R) =
\frac{1}{c^2}\int_{-r}^{r} d\sigma \frac{\epsilon_\sigma}
{\gamma_{u_\sigma}}
\delta^{(3)}\left[\vec{x}_R - \sigma\vec{n}(t_R)\right]
u^\mu(\sigma,t_R) u^\nu(\sigma,t_R)\,.
\label{t-eps-r}
\end{equation}
This expression is a sum of stress-energy tensors for point-like
particles of masses $\epsilon_\sigma d\sigma/c^2$, positions 
$\sigma\vec n(t_R)$ and four-velocities $u^\mu(\sigma,t_R)$, 
of which the string is meant to be made. The parameter 
$\sigma$ labels points on the string so that $\sigma=r$ \
corresponds to the position of the first endpoint mass,
while $\sigma=-r$ to the second one. The value of $|\sigma|$
is equal to the distance of a point from the center of the rotor.
This means, in particular, that $u_1^\mu(t_R) = u^\mu(r,t_R)$,
$u_2^\mu(t_R) = u^\mu(-r,t_R)$.

The symbol $\epsilon_\sigma$
denotes the one-dimensional, or linear, energy density of the
string at the point labeled by $\sigma$. The linear energy
density should not be confused with the energy density 
$\epsilon$ in Eq.~(\ref{tmn-fluid}), which is a three-dimensional
energy density of a medium. Nevertheless, Eq.~(\ref{t-eps-r})
gives a term in the string energy-momentum tensor that is
analogous to the first term on the right-hand side of
Eq.~(\ref{tmn-fluid}).

The symbol $\gamma_{u_\sigma}$ is the Lorentz factor
corresponding to the speed $u_\sigma = \Omega\sigma$,
with which the point labeled by $\sigma$ moves on a circle
in the frame $R$. Note that the linear energy density
$\epsilon_\sigma$ ought to include the elastic strain
energy density at the point labeled by $\sigma$ due to
the string tension at that point.

\subsubsection{ Divergence of the part of $T$ due to motion }

$T_m$ and $T_{\epsilon}$ alone are not sufficient
to ensure the laws of conservation of energy and
momentum, because Eq.~(\ref{conservation-law})
is not satisfied for them.
The four-divergence of $T_{\epsilon R}$ consists
of two terms,
\beq
\partial_0 T_{\epsilon R}^{0\nu} 
\es
\frac{1}{c^2}
\int_{-r}^{r} d\sigma \ 
\frac{\partial}{\partial ct }
\left\{
\epsilon_\sigma
\,c\, u^\nu 
\delta^{(3)}\left[\vec x - \vec r(t,\sigma)\right]
\right\} \, ,
\label{term1} 
\\ 
\label{term2}
\partial_i T_{\epsilon R}^{i\nu} 
\es
\frac{1}{c^2}
\int_{-r}^r d\sigma \
\epsilon_\sigma\, u^\nu 
\, \vec u_\sigma \cdot\vec\nabla\,
\delta^{(3)}\left[\vec x - \vec r(t,\sigma)\right] \,,
\eeq
where $\vec r(t,\sigma) = \sigma\vec n(t)$,
$\vec u_\sigma = \dot{\vec r}$, $\beta_{u_\sigma} = u_\sigma/c$,
$u_\sigma = |\vec u_\sigma|$, $u^\nu = u(\sigma,t_R)^\nu
= [\gamma_{u_\sigma} c , \gamma_{u_\sigma} \vec u_\sigma]^\nu$,
and $i=1,2,3$.
Using
\beq
\dot{\vec r}\cdot\vec\nabla\,
\delta^{(3)}\left[\vec x - \vec r(t,\sigma)\right] 
\es
-\partial_t\,
\delta^{(3)}\left[\vec x - \vec r(t,\sigma)\right] \ ,
\label{usun-pochodna-t}
\eeq
in Eq.~(\ref{term2}), and adding Eq.~(\ref{term2}) 
to Eq.~(\ref{term1}), we get
\begin{eqnarray}
\label{fourdiv-teps}
\partial_\mu T_{\epsilon R}^{\mu\nu} \es
\frac{1}{c^2}\int_{-r}^r d\sigma \ 
\delta^{(3)}\left[\vec x - \vec r(t,\sigma)\right] \
\frac{\partial}{\partial t}\left(
\epsilon_\sigma u^\nu \right)
\\
\label{fourdiv-teps1}
\es \frac{1}{c^2}\int_{-r}^r d\sigma \ 
\delta^{(3)}\left[\vec x - \vec r(t,\sigma)\right] \
\epsilon_\sigma \gamma_{u_\sigma}
\left[0,\sigma\ddot{\vec n}\right]^\nu \,.
\end{eqnarray}
Calculation of the four-divergence of $T_m$ in the
frame $R$
is done in the same steps as for $T_\epsilon$.
The result is
\begin{equation}
\partial_\mu T_{m R}^{\mu\nu} =
\delta^{(3)}[\vec x_R - \vec r_1(t_R)]
\
m \gamma_{u_ r} 
\
\left[0,r \ddot{\vec n}(t_R)\right]^\nu
+
\delta^{(3)}[\vec x_R - \vec r_2(t_R)]
\
m \gamma_{u_r} 
\
\left[0,-r \ddot{\vec n}(t_R)\right]^\nu \, .
\label{fourdiv-tm}
\end{equation}
Components zero and $z$ in both Eq.~(\ref{fourdiv-teps1})
and Eq.~(\ref{fourdiv-tm}) are equal zero, but for the
components $x$ and $y$ the sum of the two equations does
not vanish, irrespective of the shape of $\epsilon_\sigma$.
The two Dirac $\delta$-functions in Eq.~(\ref{fourdiv-tm})
are located in two different points in space -- the two ends
of the string. Their sum cannot give zero. This fact corresponds
to the paradox described in Sec.~\ref{sec:2masses}.

Addition of Eq.~(\ref{fourdiv-teps1}) is not sufficient. The points
with $\sigma=-r$ and $\sigma=r$ under the integral have measure
zero. Hence, they contribute negligible amounts to the result
of integration, in comparison to the two Dirac $\delta$-functions
in Eq.~(\ref{fourdiv-tm}). Furthermore, demanding that $\partial_\mu T_{\epsilon R}^{\mu\nu}$
in Eq.~(\ref{fourdiv-teps1}) vanishes would imply that
$\epsilon_\sigma$ vanishes for every value of $\sigma$, because
the Dirac $\delta$-functions under the integral are located in
different points in space for different values of $\sigma$.

To make four-divergence of $T_m+T_\epsilon$ vanish, we have
to put $m=0$ and $\epsilon_\sigma=0$, which would remove
the rotor energy-density entirely. This result shows that
we need another contribution to the energy-momentum of the
rotor, which is not like the contributions we can imagine
as built just from moving particles.

\subsubsection{ Parts of tensor $T$ due to stresses }

The full stress-energy tensor of the rotor has to include
fluxes of four-momentum related to forces present in the
system, {\it i.e.}, the forces responsible for the 
circular motion of the rotor. Given the interpretation 
of $T^{ij}$ components, the form of the missing part of 
the tensor can be guessed on the basis of the fact that
the momenta of the system parts always change only
along the rotating vector $\vec n$. For example, at
a moment when the string is placed along the $x$-axis,
the tension is also along the $x$-axis. In other words,
the flux of momentum which is produced by tension is
directed in the $x$ direction. It goes through the surface
perpendicular to the $x$-axis. Its only nonzero component
is the $x$ component. Therefore, only $T^{xx}$ component 
of the energy-momentum tensor is nonzero at times when 
$n_x=1$ and $n_y=0$. When the string is placed along
$y$-axis, and $n_x=0$, $n_y=1$, then only $T^{yy}$ component
is nonzero. At other positions, the stress is along $\vec n$.
Thus, our guess is that $T^{ij}\sim n_i n_j$.
Therefore, we write
\begin{equation}
T_{p R}^{\mu\nu}(t_R,\vec x_R) =
\int_{-r}^r d\sigma \frac{p_\sigma}{\gamma_{u_\sigma}}
\delta^{(3)}\left[\vec x_R - \sigma\vec n(t_R)\right]
\left[0,\vec n(t_R)\right]^\mu
\left[0,\vec n(t_R)\right]^\nu\,,
\label{t-p-r}
\end{equation}
where $p_\sigma$ is the tension of the string in the
local rest frame of reference of the string piece 
labeled by $\sigma$, called below the $\sigma$-piece 
of the string. Note that the linear tension $p_\sigma$
in Eq.~(\ref{t-p-r}) differs from the three-dimensional 
pressure $p$ in Eq.~(\ref{tmn-fluid}), although
the two stress-energy tensors play similar physical 
roles. 

Equation (\ref{t-p-r}) gives the term in the stress-energy 
tensor of the string analogous to the second term on 
the right-hand side of Eq.~(\ref{tmn-fluid}). The analogy
appears because $-\left[0,\vec n(t_R)\right]^\mu
\left[0,\vec n(t_R) \right]_\nu$ is the projection operator
on the spacetime  direction of $\left[0,\vec n(t_R)\right]^\mu$.
One can say that the string $T_{p}$ in Eq.~(\ref{t-p-r})
is similar to the perfect fluid tensor $T$ of Eq.~(\ref{tmn-fluid})
in the sense that in the string the momentum exchange between
its parts happens in one direction, along the string, and in
the fluid the momentum is exchanged between elements of the
fluid in all spatial directions.

\subsubsection{ Divergenceless energy-momentum tensor }

The sum of (\ref{t-m-r}), (\ref{t-eps-r})
and (\ref{t-p-r}) is sufficient to define the rotor
energy-momentum tensor with vanishing four-divergence.
The four-divergence of $T_p$ is
\begin{align}
\partial_\mu T_{pR}^{\mu\nu}(t,\vec x) = \int_{-r}^r d\sigma
\frac{p_\sigma}{\gamma_{u_\sigma}} \left[0,\vec n(t)\right]^\nu
\vec r\,' \cdot \vec\nabla
\delta^{(3)}\left[\vec x - \vec r(t,\sigma)\right]\,,
\label{fourdiv-tp}
\end{align}
where prime denotes differentiation with respect 
to $\sigma$ and we replaced $\vec n$ with $\vec r\,'
=(\sigma\vec n)'= \vec n$.
Using the identity 
\beq
\vec r\,'\cdot\vec\nabla\,\delta^{(3)}
\left[\vec x - \vec r(t,\sigma)\right] 
\es 
-\partial_\sigma\,
\delta^{(3)}\left[\vec x - \vec r(t,\sigma)\right]
\label{usun-pochodna-s}
\eeq
and integrating by parts over $\sigma$, we remove the
differentiation from the Dirac $\delta$-function and
move it to $p/\gamma$. The components $\nu=0$ and $\nu=z$
in Eq.~(\ref{fourdiv-tp}) vanish while the components
$\nu=x$ and $\nu=y$ are proportional to $n_x$ and $n_y$,
respectively. This procedure yields also the boundary
terms proportional to three dimensional $\delta$-functions
$\delta^{(3)}[\vec x_R - \vec r_1(t_R)]$ and
$\delta^{(3)}[\vec x_R - \vec r_2(t_R)]$,
corresponding to those present in Eq.~(\ref{fourdiv-tm}).

We now recall that $\partial_\mu T_{\epsilon R}^{\mu i}$
is proportional to $\ddot{r}^i$, which is proportional
to $\vec n$. Thus, given the forms of $T_\epsilon$
and $T_p$, we can adjust $\epsilon_\sigma$ and $p_\sigma$
to make the four-divergence of 
\beq
\label{fullT}
T \es T_m + T_\epsilon + T_p
\eeq
vanish. The four-divergence of $T$ is given by sum of
Eqs.~(\ref{fourdiv-teps1}), (\ref{fourdiv-tm}) and
(\ref{fourdiv-tp}). Factors in front of respective
Dirac $\delta$-functions have to cancel for every
$\sigma$ and in the endpoints separately.
Equation~(\ref{fourdiv-teps1}) and string part of
Eq.~(\ref{fourdiv-tp}) give
\beq
\frac{d}{d\sigma} \ \frac{p_\sigma} {\gamma_{u_\sigma}} 
\es
\gamma_{u_\sigma} \ \frac{\epsilon_\sigma}{c^2}
\
\Omega^2\sigma
\, ,
\label{dynamic-eqn} 
\eeq
where $\Omega^2$ comes from $-\ddot{\vec n}$.
Equation~(\ref{fourdiv-tm}) and boundary terms
from Eq.~(\ref{fourdiv-tp}) result in the boundary condition
\beq
\frac{p_r}{\gamma_{u_r}} 
\es
\frac{p_{-r}}{\gamma_{u_r}} 
\rs
- \ \gamma_{u_r} \ m \ \Omega^2 r \, .
\label{boundary-condition}
\eeq
These equations show that if we want the total four-momentum of
the rotor to be conserved in every frame of reference,
{\it i.e.}, if we want Eq.~(\ref{conservation-law}) to
hold, then the energy-momentum tensor of the rotor has to
include the part in Eq.~(\ref{t-p-r}). This part describes
the stresses in the string that are responsible for what
we might call the forces that bind the system. The string
tension resulting from these stresses causes the endpoint
masses to move on a circle in the frame $R$. However, the
concept of force is not a natural or intuitive one here
because the string may have complex physical properties.
They may result in complex solutions to Eq.~(\ref{dynamic-eqn}) 
with boundary conditions of Eq.~(\ref{boundary-condition})
for the tension $p_\sigma$ and linear energy density
$\epsilon_\sigma$ as functions of $\sigma$.

In Sec.~\ref{sec:eqn-analysis} we analyze Eqs.~(\ref{dynamic-eqn})
and (\ref{boundary-condition}), which ensure Eq.~(\ref{conservation-law}).
Subsequently, in Secs.~\ref{sec:en-mom-moving-frame},
\ref{sec:examples} and \ref{sec:ultrarelativistic}, we study
the contribution of $T_p$ to the rotor total 
four-momentum in the frame $M$, including examples.

\subsection{ Analysis and interpretation of
             Eqs.~(\ref{dynamic-eqn}) and
             (\ref{boundary-condition}) }
\label{sec:eqn-analysis}

Let us first analyze Eq.~(\ref{boundary-condition}).
The factor $m \, \Omega^2r = m \, u_r^2/r$ on the right hand side, which
is equal to the usual, nonrelativistic centripetal force,
is needed to make a point mass move on a circle. The
relativistic correction, provided by the factor
$\gamma_{u_r}$, has its origin in the relativistic
definition of momentum $\vec P = \gamma_v \,  m \vec v$
of a particle with mass $m$ and velocity $\vec v$.
A change in a particle's momentum is a result of action
of a force, which quantitatively is described by
equation $\vec F = \dot{\vec P}$. In uniform circular motion,
the magnitude of velocity is constant, $u_r = \Omega r$, and
the magnitude of centripetal force is $\gamma_{u_r} \, m u_r^2 /r$.
Hence, the right hand side of Eq.~(\ref{boundary-condition})
is, up to a sign, equal to the centripetal force needed to keep
the endpoint masses moving on a circle in the frame $R$.
The left-hand side of Eq.~(\ref{boundary-condition}) comes from 
the string, the extended system feature which provides
the centripetal force that binds the endpoint masses.

We conclude that in the frame $R$, the magnitudes of forces
provided by the string that act on the endpoint masses are 
$p_r/\gamma_{u_r}$ and $p_{-r}/\gamma_{u_r}$.
Furthermore, the tension of the string at point $\sigma$ as
seen in the frame $R$ is $p_\sigma/\gamma_{u_\sigma}$.
This conclusion agrees with our interpretation of
$p_\sigma$ as the tension of the $\sigma$-element of the
string in its local rest frame. If in the local rest frame of
$\sigma$-element, say for $\sigma=r$, the second law 
of dynamics has the form $\vec F = d\vec P/d\tau$, where $\tau$ 
is the time parameter in that frame, then in the frame $R$
it will have the form $\vec F = \gamma d\vec P/dt_R$, where
$\gamma=dt_R/d\tau$. We assume that $\vec F$ and $d\vec P$
do not change between these two frames of reference, because
the string element moves in $R$ perpendicularly to its axis.
So, while  $\vec F$ is the centripetal force in the local
rest frame, in the frame $R$ the centripetal force is equal
$\vec F/\gamma$.

The content of Eq.~(\ref{dynamic-eqn}) can be studied
in the nonrelativistic limit ($c\to\infty$), since it
simplifies in that limit and the meaning of the gamma
factors is already explained in our analysis of
Eq.~(\ref{boundary-condition}). Assuming $\epsilon=\rho c^2$,
where $\rho$ is the linear mass density on the string,
the nonrelativistic limit of Eq.~(\ref{dynamic-eqn}) is
\beq
\frac{d p}{d\sigma} 
\es
\rho \ \Omega^2 \sigma  
\rs
\rho \ u_\sigma^2 / \sigma  
\, .
\label{dynamic-eqn-NR}
\eeq
This equation describes an infinitesimal
part of the string, which is located at the point labeled
by $\sigma$ and has length $d\sigma$. Thus, $dp$ is the 
radial component of the force acting on the considered 
string part. One end is pulled inwards with force $p$, 
the other end is pulled outwards with force $p+dp$. 
Components other than radial are zero. The resultant 
force equals the centripetal force making the element 
move on a circle, $\rho d\sigma \ u_\sigma^2/\sigma$.
This is the content of Eq.~(\ref{dynamic-eqn-NR}).

One implication of Eq.~(\ref{dynamic-eqn}) and
Eq.~(\ref{boundary-condition}) is that the tension $p$
is negative. It follows from the fact that
the boundary tension $p_r$ is negative and $p_\sigma/\gamma_{u_\sigma}$
is decreasing for $\sigma<0$ and increasing for
$\sigma>0$. This fact can be understood if we
imagine the string as a pipe filled with a fluid. 
A string at rest corresponds to pressure zero. 
Shortening the pipe results in a positive and
extending the pipe results in a negative pressure
inside the pipe. A rotating string is stretched to
provide the centripetal force to the rotor massive
endpoints and every other point of the string.
Hence $p$ is negative for the stretched string.

\subsubsection{ Solution for the tension
                   $p_\sigma$ in Eq.~(\ref{dynamic-eqn}) }

We present the solution of Eq.~(\ref{dynamic-eqn})
assuming that the string energy density $\epsilon_\sigma$ 
is given {\it a priori}, and that as such it satisfies 
the symmetry condition $\epsilon_{-\sigma} = \epsilon_\sigma$. 
Equation~(\ref{dynamic-eqn}) can be written in the form
\beq
\frac{d}{d\sigma}
\
\frac{p_\sigma}{\gamma_{u_\sigma}}
\es
-\left(\frac{1}{\gamma_{u_\sigma}}\right)'
\epsilon_\sigma\,,
\label{qprim}
\eeq
where the prime means differentiation with respect
to $\sigma$. This form suggests that integration
on the right-hand side could be done by changing
the integration variable $\sigma$ to the variable
$g = 1/\gamma_{u_\sigma}$. Consequently, the solution
to Eq.~(\ref{qprim}) can be written in the form
\beq
\frac{p_\sigma}{\gamma_{u_\sigma}} \es p_0
+ \int_{1/\gamma_{u_\sigma}}^1 \epsilon_{\sigma_g} dg
\, ,
\label{solution-p}
\eeq
where $\sigma_g =\tfrac{c}{\Omega}\sqrt{1-g^2}$ and $p_0$
is the string tension in the middle, {\it i.e.},
at $\sigma=0$. The solution is valid in 
the whole range of $\sigma$ from $-r$ to $r$.

\subsection{ String contribution to the rotor four-momentum
in the moving frame }
\label{sec:en-mom-moving-frame}

Having secured the energy and momentum conservation via
four-dimensionally divergenceless energy-momentum tensor,
we can now analyze the contribution of the string to the
rotor four-momentum in the moving frame $M$. The stress-energy
tensor of Eq.~(\ref{fullT}) in that frame is given by
\begin{equation}
 T_{M}^{\mu\nu}(t_M,\vec x_M) =
 L^\mu_{\,\,\alpha} \, L^\nu_{\,\,\beta} \,
 T_{R}^{\alpha\beta}\left[  t_R(t_M, \vec x_M),
 \vec x_R(t_M, \vec x_M) \right]\,,
 \label{tmn-in-M}
\end{equation}
where $L^\mu_{\,\,\alpha}$ is the Lorentz transformation
matrix for the boost from the frame $R$ to the frame $M$,
and $t_R$, $\vec x_R$ are functions of $t_M$, $\vec x_M$
that are determined by the Lorentz transformation $L^{-1}$. 
Integrating the zero-component of the tensor density over 
space in the frame $M$ according to Eq.~(\ref{Pt}), one 
obtains the following example of a rotating extended
system four-momentum density in motion in the form
\beq
\frac{dP_M}{d\sigma} \es
\frac{1}{c}\epsilon_\sigma\gamma_\sigma
\begin{bmatrix}
\gamma_U \\
\beta_U \gamma_U \\
0 \\
0
\end{bmatrix}
+
\frac{1}{c}\epsilon_\sigma\gamma_\sigma
\beta_\sigma 
\begin{bmatrix}
 -\beta_U \gamma_U s_\sigma \\
 -        \gamma_U s_\sigma \\
                   c_\sigma \\
 0
\end{bmatrix}
- \frac{p_\sigma }{c \gamma_\sigma}
\frac{ \beta_U c_\sigma}
{1 - \beta_U \beta_\sigma s_\sigma}
\begin{bmatrix}
 -\beta_U \gamma_U c_\sigma \\
 -        \gamma_U c_\sigma \\
 -                 s_\sigma \\
 0
\end{bmatrix} \ ,
\label{momentum-density}
\eeq
where $u_\sigma = \Omega \sigma$, $\beta_\sigma = u_\sigma/c$
and $\gamma_\sigma = (1-\beta_\sigma^2)^{-1/2}$. The first two
terms come from the term $T_\epsilon$
and the third one comes from $T_p$ 
in Eq.~(\ref{fullT}), while 
\beq
s_\sigma \es \sin[\Omega t_R(t_M,\sigma)] \ , \\
c_\sigma \es \cos[\Omega t_R(t_M,\sigma)] \ ,
\eeq
denote the trigonometric factors. Here, $t_R(t_M,\sigma)$
is a generalization of functions $t_{1R}(t_M)=t_{R}(t_M,r)$
and $t_{2R}(t_M) = t_{R}(t_M,-r)$, obtained by inserting
$\sigma\cos\Omega t_R$ in place of $x_R$ in
Eq.~(\ref{lorentz-time}), and solving for $t_R$.

The terms originating in the part $T_\epsilon$ hold no surprises: 
the energy and momentum of a $\sigma$-piece of the string 
are its Lorentz transformed energy and momentum from the 
rotor rest frame $R$. Similarly, as observed in
Sec.~\ref{sec:2masses}, the total energy and momentum resulting
from the part $T_m$ are not constant in time in the frame $M$.
The same holds true for the total energy and momentum that come
from integrating the sum $T_m+T_\epsilon$. In these parts, a
piece of a string corresponding to some value of $\sigma$
resembles a particle with mass $\epsilon_\sigma d\sigma/c^2$.

In contrast, the part $T_p$ only contributes to the 
energy-momentum of the rotor in the frame $M$. It does 
not contribute anything to the rotor energy and momentum 
in the frame $R$.
The tensor structure of $T_p$ in the frame $M$ is given
by $w_M^\mu w_M^\nu$, where $w_M^\mu$ is a four-vector
obtained by applying the Lorentz transformation to
$w_R^\mu = [0,\vec n(t_R)]^\mu$, so that
\begin{equation}
 w_M^\mu = L^\mu_{\,\,\alpha} w_R^\alpha =
 \left[\begin{array}{c}
  \beta_U\gamma_U\cos\Omega t_R\\
  \gamma_U\cos\Omega t_R\\
  \sin\Omega t_R\\
  0
 \end{array}
 \right]^\mu\,.
\end{equation}
Leaving out the factors in front and the Dirac
$\delta$-functions, we can identify the form of
``tension'' in stress-energy tensor of $\sigma$-element
using the matrix,
\begin{equation}
 T_{pM}^{\mu\nu}(\sigma\text{-element}) \sim p_\sigma
 \left[\begin{array}{cccc}
  \beta_U^2 \gamma_U^2 c_\sigma^2 & \beta_U \gamma_U^2 c_\sigma^2 & \beta_U \gamma_U s_\sigma c_\sigma & 0\\
  \beta_U   \gamma_U^2 c_\sigma^2 &         \gamma_U^2 c_\sigma^2 &         \gamma_U s_\sigma c_\sigma & 0\\
  \beta_U   \gamma_U   s_\sigma c_\sigma &  \gamma_U s_\sigma c_\sigma &  s_\sigma^2 & 0\\
  0 & 0 & 0 & 0
 \end{array}
 \right]^{\mu\nu}\,,
 \label{tpMmunu}
\end{equation}
where again $s_\sigma = \sin\Omega t_R(t_M,\sigma)$ and
$c_\sigma = \cos\Omega t_R(t_M,\sigma)$.

Since the zero-zero component of the energy-momentum tensor equals
the energy density, the energy of a string $\sigma$-element
is  proportional to $p_\sigma\beta_U^2\gamma_U^2\cos^2\Omega t_R$.
Since $p$ is negative, the part of energy of $\sigma$-element
that results from the tension is also negative. Similarly,
the $x$ component of momentum of a $\sigma$-element is proportional
to $p_\sigma\beta_U\gamma_U^2\cos^2\Omega t_R$. For $\beta_U$ positive,
the rotor as a whole has positive $x$ component of momentum,
but the momentum carried by a $\sigma$-element has a negative 
$x$ component. For $\beta_U$ negative, the rotor velocity is
negative and the momentum of a $\sigma$-element is positive
in the $x$ direction. Additionally, we have the $y$ component
of the momentum of a $\sigma$-element, which does not have
a definite sign. This indicates that the string momentum part
resulting from the tension is approximately antiparallel
to the velocity of the rotor as a whole. The described 
features can be observed in the examples presented in Sec.~\ref{sec:examples},
see Figs.~\ref{fig:momentum} and \ref{fig:momentum-hooke}.

The energy and momentum carried by the string have different
properties from the ones exhibited by point-like massive particles.
For example, a particle carries energy as a zero component 
of a four-vector and its energy is a non-zero quantity in all
frames of reference. In contrast, the string tension contributes
to energy as a zero-zero component of a tensor. It is possible
that the tension contributes non-zero energy in the moving frame
$M$ even if in some frame of reference, such as the rest frame $R$,
it contributes no energy. The same happens with momentum.

Our analysis of the rotor indicates that similar intricacies 
can be expected in relativistic description of dynamics of all 
extended physical systems in motion, irrespective of the internal 
mechanisms that bind them. In the case of a rotor, the string 
contribution to the total energy and momentum in the moving 
frame, resulting from the string tension, exists regardless of
the specific string properties that lead to concrete functions 
$\epsilon_\sigma$ and $p_\sigma$. Illustrative examples of these 
functions are presented in Sec.~\ref{sec:examples}.

\subsection{ Examples of the string }
\label{sec:examples}

We consider two examples of string behavior.
Our first, simple model assumes that the string 
linear energy density is constant as a function of 
the parameter $\sigma$. Our second model 
concerns the string that stretches according
to the Hooke law.

\subsubsection{ Simple string }
\label{sec:simple-string}

In our first example of the string, we assume the linear
energy density $\epsilon_\sigma = \rho c^2 = \text{const}$.
Then, using the solution of Eq.~(\ref{dynamic-eqn}), given
in Eq.~(\ref{solution-p}), we have
\beq
\frac{p_\sigma}{\gamma_\sigma} \es
p_0 + \rho c^2 \left( 1-\frac{1}{\gamma_\sigma} \right) \, ,
\label{sol-muconst}
\eeq
where $\beta_\sigma = \Omega \sigma/c$.
From the boundary condition, Eq.~(\ref{boundary-condition}),
we can determine $p_0$, which is the tension
of the string in the center of the rotor,
\beq
p_0 \es - \gamma_r m \, \Omega^2 r
- \rho c^2\left[1 - \frac{1}{\gamma_r}\right]
\,.
\label{muconst-p0}
\eeq
Now, putting this $p_0$ into Eq.~(\ref{sol-muconst}),
we arrive at
\beq
\frac{p_\sigma}{\gamma_\sigma} 
\es
- \gamma_r m \, \Omega^2 r
- \rho c^2 \left[ \frac{1}{\gamma_\sigma}
-\frac{1}{\gamma_r} \right] \ ,
\eeq
or, alternatively written,
\beq
\frac{p_\sigma}{\gamma_\sigma} \es
- \gamma_r \, m \, u_r^2/r
- \frac{\rho(u_r^2-u_\sigma^2)}{1/\gamma_r
+ 1/\gamma_\sigma} \,.
\label{solution-p-mu-const}
\eeq
If $\Omega r$ is small compared to the speed
of light, the solution may be approximated by
\begin{equation}
p_\sigma = - \, \frac{ m u_r^2 }{ r}
-\frac{\rho (u_r^2-u_\sigma^2)}{2}
+ O(\beta_r^2 )\,.
\label{p-mu-const-nonrel}
\end{equation}
This nonrelativistic limit has the following interpretation:
the string tension at $\sigma$, denoted by $p_\sigma$, is 
the sum of centripetal forces needed to hold the endpoint 
mass and outer part of the string beyond $\sigma$
in their circular motion. The first term, equal $-m\Omega^2r$,
corresponds to the endpoint mass. The second term is the 
sum of centripetal forces corresponding to all infinitesimal 
pieces of the outer part of the string or, effectively, of 
a body of mass $(r-\sigma)\rho$ rotating with frequency 
$\Omega$ at the distance $(r+\sigma)/2$ from the string 
center. 

The relativistic result of Eq.~(\ref{solution-p-mu-const})
is more complicated than its nonrelativistic limit.
In Sec.~\ref{sec:string-stress-energy-tensor} we argue
that the tension as seen in the frame $R$ equals
$p_\sigma/\gamma_\sigma$. Since the first term
on the right hand side of Eq.~(\ref{solution-p-mu-const})
is equal to the centripetal force acting on the endpoint
mass, including the gamma factor $\gamma_r$, the second
term can be interpreted as the centripetal force which
holds the outer part of the string or effectively the
force needed to hold one body of mass $(r-\sigma)\rho$ at
position $(r+\sigma)/2$, with effective gamma factor being
the harmonic mean of $\gamma_r$ and $\gamma_\sigma$.

\begin{figure}[!h]
\includegraphics{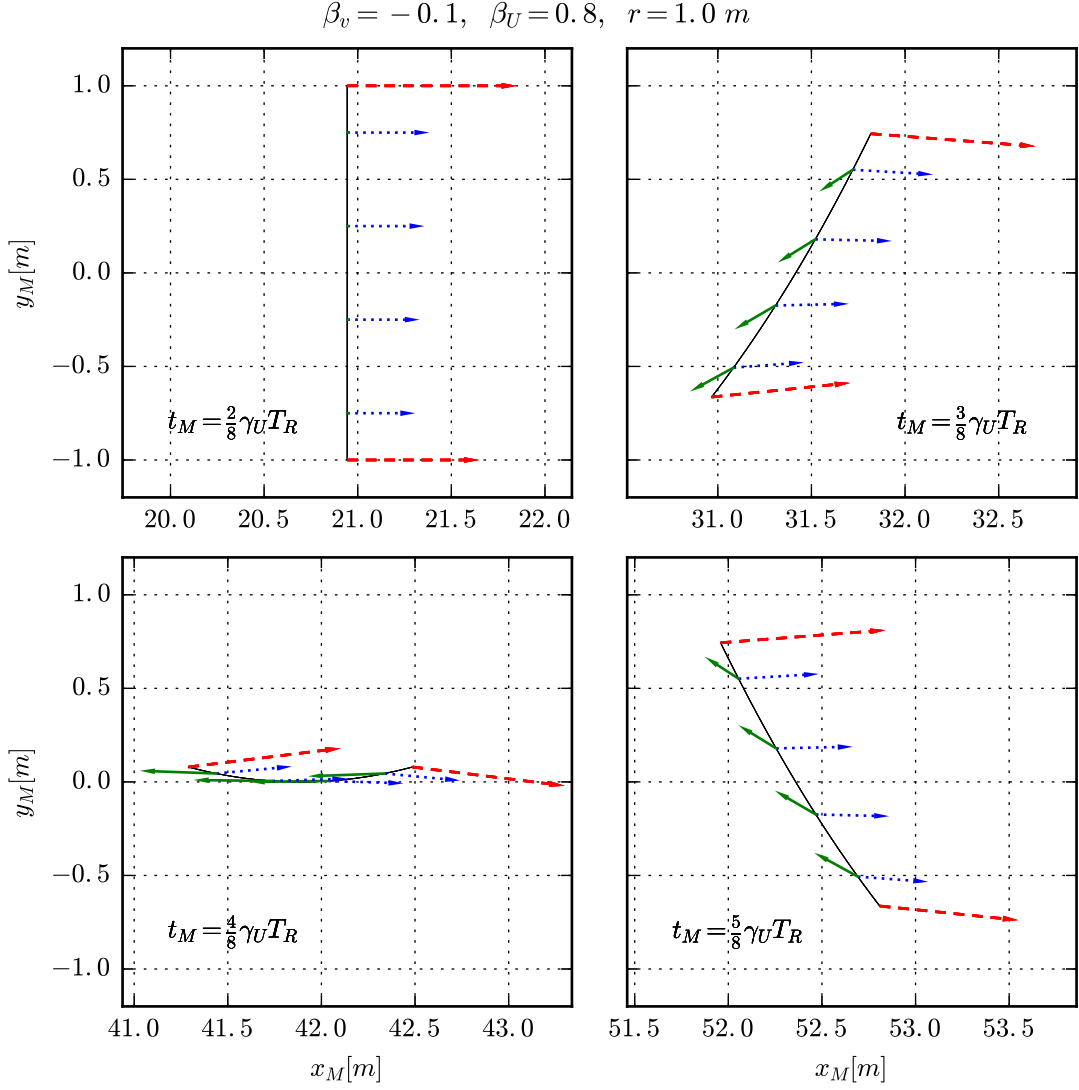}
\caption{
	Momentum carried by different parts of the rotor at four
	different times in the moving frame. Here, $T_R$ means the
	period of rotation in the frame $R$. The string is drawn
	with a solid line (black in the online version). Dashed arrows
	(red in the online version) indicate momenta of endpoints.
	Dotted arrows (blue in the online version) indicate the
	density of momentum of the string which comes from
	$T_\epsilon$ at four points on the string, for $\sigma/r$ =
	$-0.75$, $-0.25$, $0.25$, $0.75$. Similarly, solid arrows
	(green in the online version) indicate the density of momentum
	which comes from $T_p$. The string is assumed light and the
	rotation of the rotor is not very fast. Therefore, the dotted
	and solid arrows are magnified 10 and 100 times with respect
        to the dashed ones, correspondingly. We assume the simple model
	of the string from Sec.~\ref{sec:simple-string}.
}
\label{fig:momentum}
\end{figure}

An example of our rotor is presented in Fig.~\ref{fig:momentum}.
The figure shows momentum carried by different parts of the rotor 
as observed in the moving frame, assuming the simple model of the 
string with $\rho = 0.1$~kg/m. The rotor radius $r$ is $1$~m and 
the mass of each endpoint is $1$~kg. The momentum density on the 
string is calculated using Eq.~(\ref{momentum-density}). The result 
is divided into the part proportional to $\epsilon$ (dotted arrows) 
and the part proportional to $p$ (solid arrows). These parts 
correspond to the parts $T_\epsilon$ and $T_p$ of the total 
energy-momentum tensor, respectively. Momenta carried by the endpoints,
which correspond to $T_m$, are indicated by the dashed arrows.
One cannot directly compare momentum with momentum density
because they have different meanings and dimensions. Therefore, 
we multiplied the densities by $r/2$, which is one fourth of the 
total length of the string. Thus, the arrows presented in the figure 
are approximately equal to momenta carried by pieces of the string of 
length $r/2$. Furthermore, because the string is much lighter than 
the endpoint masses, the dotted arrows are magnified 10 times with 
respect to the dashed ones and, because $p_0/(\rho c^2) \approx -0.11$, 
the solid arrows are magnified 100 times with respect to
the dashed ones. Otherwise, dotted and solid arrows would
not be well visible. The frame $M$ moves with respect to
the frame $R$ with the speed $U=0.8c$. The speed of each
endpoint in the frame $R$ is $v=-0.1c$, where the minus means
that the rotor rotates clockwise when looked at from $z>0$
half-space. However small, $T_p$ contributes to the total
momentum of the rotor and its contribution is different
from that of $T_\epsilon$. For example, the solid arrows do
not have the direction of velocity of the corresponding points
of the string, contrary to the dotted arrows that represent
the momentum density coming from $T_\epsilon$. Another visible
aspect of the rotor motion is the apparent bending of the string 
in the frame $M$. It occurs even though the string is straight
at every moment in the frame $R$. This is a consequence of 
relativity of simultaneity, {\it cf.} Fig. 7 in Ref.~\onlinecite{trolley}.

\subsubsection{ Hooke's string }
\label{sec:hooke}

In the previous example, we assume that the string linear
energy density $\epsilon$ is constant and we use this constant
to derive the tension $p$. But in general, $\epsilon$ will
depend on $p$. Both quantities 
depend on how much the string is stretched. If $p_\sigma$
is uniquely determined by the elongation of the element
of the string labeled by $\sigma$, then $\epsilon$ can
be considered a function of $p$. Therefore, we do not
need to know $\epsilon$ in advance, if we know the law
governing the stretching of the string.

To properly account for the effects of stretching of the
string, we need to define a quantity which encodes information
about its deformation. Assuming that the energy does not
depend on whether the string is bent or straight, we need
only one number to characterize the degree of stretching
locally on the string, and we denote it by $k_\sigma$.
We define $k_\sigma$ as the ratio of length of the infinitesimal
string element that is relaxed, to the actual length of the
same element of the string when it is stretched or compressed.
If $k_\sigma$ is one, then the piece of the string labeled by 
$\sigma$ is relaxed. If $k_\sigma<1$, then the element is stretched. 
When $k_\sigma>1$, then the element is compressed. We will call 
$k_\sigma$ the strain parameter, corresponding to the string 
parameter $\sigma$.

We now assume that the string obeys Hooke's law.
For a moment, let us ignore the parameter $\sigma$ and
let's assume we have a piece of string at rest, which 
is stretched from length $l_0$ to $l$. The tension $p$ 
is
\begin{equation}
p = -\frac{Y}{l_0}(l-l_0)
= -Y \left( \frac{1}{k}-1 \right)\,,
\label{p-hook}
\end{equation}
where $k=l_0/l$ and $Y$ is a material constant of
dimension of force; for a string 
of a negligibly small but finite thickness it would be
the Young modulus times the cross-section area of 
the string. The energy needed to stretch the string 
is $\frac{Y}{2l_0} (l-l_0)^2$. If the string has 
energy $\varepsilon_0 l_0$ before stretching, where
$\varepsilon_0$ is the linear energy density of the
string in the relaxed state, then after stretching
the linear density $\epsilon$ is
\beq
\epsilon = \frac{\varepsilon_0 l_0 + 
\frac{Y}{2l_0}(l-l_0)^2}{l}
= \left(\varepsilon_0 + 
\frac{Y}{2}\right)k +\frac{Y}{2k} - Y\,.
\label{e-hook}
\eeq
Since $p$ is determined by $k$ according to
Eq.~(\ref{p-hook}), one can find the inverse
function $k(p)$, and put it into $\epsilon(k)$,
which yields
\beq
\epsilon = \frac{1}{1-\frac{p}{Y}}
\left(\varepsilon_0 + \frac{p^2}{2Y}\right)\,.
\label{eps-p}
\eeq
This relation does not depend on the length of a piece
of the string and may be considered a local property
of the string. It does not depend on the state of motion
of the string either, because $\epsilon$ and $p$ are
quantities defined point-wise in the local rest frames
of the elements of the string. These features are
also shared by Eq.~(\ref{p-hook}) and Eq.~(\ref{e-hook}). 

Using Eq.~(\ref{eps-p}), we can get rid of $\epsilon$
in Eq.~(\ref{dynamic-eqn}), which can then be solved
for $p_\sigma$. There is also another possibility,
which we pursue: one can insert Eqs.~(\ref{p-hook})
and (\ref{e-hook}) into Eq.~(\ref{dynamic-eqn}),
obtaining an equation for $k_\sigma$, which can be solved
by method of separation of variables. The result is
\beq
k_\sigma 
\es
 \frac{1}{\sqrt{B^2
-\frac{\gamma_\sigma}{\gamma_r} C }}\,,
\label{hook-k}
\eeq
where
\beq
B^2 \es 1+\frac{2\varepsilon_0}{Y}\,,\\
C \es B^2 - \left[1 + \gamma_r^2 \frac{m u_r^2
}{r Y }\right]^2\,.
\eeq
Using this result, $p_\sigma$ and $\epsilon_\sigma$ 
can be obtained by inserting $k_\sigma$ 
of Eq.~(\ref{hook-k}) into Eq.~(\ref{p-hook}) 
and Eq.~(\ref{e-hook}), respectively.

This solution does not present itself as intuitively
understandable, contrary to Eqs.~(\ref{solution-p-mu-const})
and (\ref{p-mu-const-nonrel}). This is because
$\epsilon$ and $p$ and the boundary conditions are
mutually related. However, in the limit of an infinitely
rigid string, that is, $Y\to\infty$, $k$ equals one,
$\epsilon \to\varepsilon_0$, and $p_\sigma$ reproduces
the solution given by Eq.~(\ref{solution-p-mu-const})
with $\rho = \rho_0 := \varepsilon_0/c^2$. One can also
study the nonrelativistic limit of $c\to\infty$.
Tension $p$ in this limit is
\begin{equation}
p_\sigma = -Y \left[ \sqrt{
\frac{\rho_0( u_r^2-u_\sigma^2)}{Y}
+\left(1 + \frac{m u_r^2 }{r Y}\right)^2
}-1 \right] + O\left(\frac{1}{c^2}\right)\,.
\label{sol-p-hook-nonrel}
\end{equation}
If we further assume that the string is very rigid
($Y\to\infty$), then the above equation assumes the
form of Eq.~(\ref{p-mu-const-nonrel}) with $\rho=
\rho_0$. 

Our conclusion is that the example of Hooke's string,
which is intended to provide more accurate description
of the string than the previous example, includes
the previous example as a special, limiting, case.
In particular, the example of a rotor depicted in
Fig.~\ref{fig:momentum} can be reproduced within few
percent accuracy with Hooke's string whose parameters
are $\varepsilon_0 = 88~\text{kg/m} \cdot c^2$ and
$Y=10^{12}$~N. The elongation of the string in that
case equals about 1000, which is far beyond the
elasticity of ordinary strings or ropes.

\begin{figure}[!h]
\includegraphics[trim={0 0 0 -0.56cm},clip]{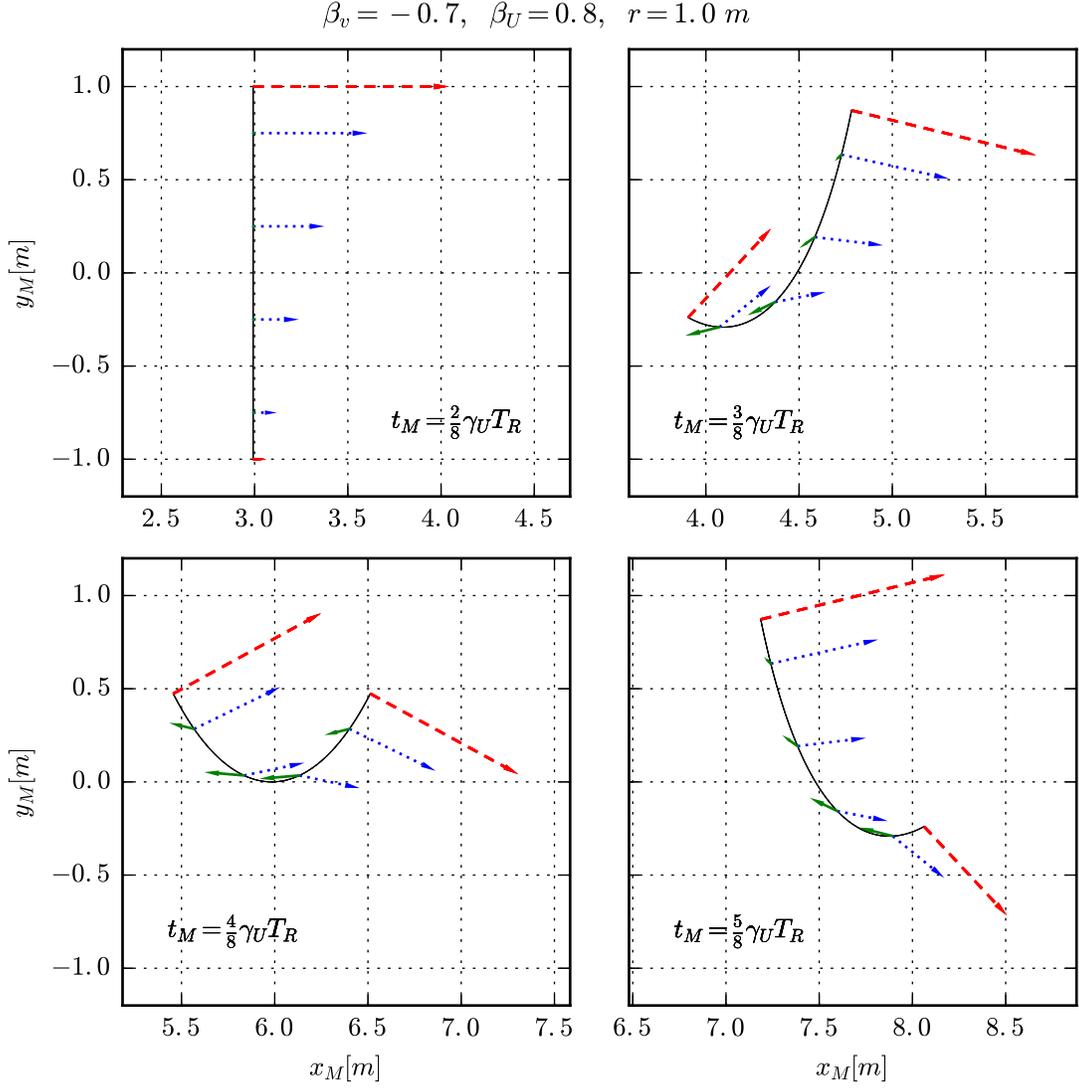}
\caption{
	Momentum carried by different parts of the rotor at four
	different times in the moving frame. $T_R$ means here the
	period of rotation in the frame $R$. The string is drawn
	with a solid line (black in the online version). Dashed arrows
	(red in the online version) indicate momenta of endpoints.
	Dotted arrows (blue in the online version) indicate the
	density of momentum of the string which comes from $T_\epsilon$
	at four points on the string, for $\sigma/r$ = $-0.75$, $-0.25$,
	$0.25$, $0.75$. Similarly, solid arrows (green in the online
	version) indicate the density of momentum which comes from
	$T_p$ in the same four points on the string. We assume that
	the string obeys Hooke's law. For more details
	see Sec.~\ref{sec:hooke}.
}
\label{fig:momentum-hooke}
\end{figure}

Another example of the rotor is presented in Fig.~\ref{fig:momentum-hooke}.
It shows momentum carried by the rotor parts in the
moving frame when the string obeys Hooke's law.
The momentum density on the string is calculated using
Eq.~(\ref{momentum-density}), where the string
energy density $\epsilon_\sigma$ and tension $p_\sigma$
are given by Eqs.~(\ref{e-hook}) and (\ref{p-hook}),
respectively, with $k$ replaced in these equations by
$k_\sigma$ of Eq.~(\ref{hook-k}).

The total momentum density is divided into the part 
proportional to $\epsilon$ (dotted arrows) and the part
proportional to $p$ (solid arrows). This division corresponds 
to the division of the total energy-momentum tensor into 
the parts $T_\epsilon$ and $T_p$, respectively. Momenta 
carried by the endpoints, which correspond to $T_m$, are 
indicated by dashed arrows. For the purpose of comparing 
quantities with different dimensions, the densities are 
multiplied by $r/2$. Therefore, every dotted and solid 
arrow might be interpreted as momentum carried by a piece 
of string of length $r/2$, measured in the frame $R$.
The string parameters are $\varepsilon_0 = 10^5~\text{kg/m}
\cdot c^2$ and $Y=10^{12}$~N. The frame $M$ moves with respect
to the frame $R$ with the speed $U=0.8c$, the speed of each
endpoint in the rotor frame $R$ is $v=-0.7c$. The rotor radius
$r$ is $1$~m and the mass of each endpoint is $1$~kg. These
values of parameters imply that the tension is comparable
with the energy density, $p_0/\epsilon_0 \approx -0.74$.
This means that the presented string is in the regime of
ultrarelativistic limit discussed in Sec.~\ref{sec:ultrarelativistic}.

Furthermore, the elongation of the string in the middle
is $1/k_0 \approx 10^5$. It is extremely large, far beyond
what ordinary strings can withstand. In other words,
the string has relaxed length of order $10$~$\mu$m and
it is stretched to $1$~m. We assumed also that the string
is extremely dense in the relaxed state, $\varepsilon_0 =
10^5~\text{kg/m}\cdot c^2$. If the string was much lighter,
then we would have $\epsilon+p<0$, which appears unphysical,
see Sec.~\ref{sec:ultrarelativistic}. The energy density
in the middle of the stretched string is $\epsilon_0 \approx
1.55~\text{kg/m} \cdot c^2$. Similarly to Fig.~\ref{fig:momentum},
solid arrows in Fig.~\ref{fig:momentum-hooke} are approximately
antiparallel to the dotted ones. The latter have the direction
of the velocity of corresponding string points, but this time
the arrows are directly comparable. The string also seems to
bend in the moving frame, although it is always straight in
the rest  frame, which can again be compared with Fig. 7
in Ref.~\onlinecite{trolley}. The effect of apparent bending
of the string is purely kinematic and much more visible here
when compared with Fig.~\ref{fig:momentum} because here we
assume much faster rotation of the rotor.

\subsection{ Ultrarelativistic limit }  %
\label{sec:ultrarelativistic}           %

Although a truly relativistic rotor cannot be built
from real atoms because the interatomic binding 
forces are too weak to withhold relativistic tension, 
it is, nevertheless, interesting theoretically what 
could happen if the angular velocity $\Omega$ were 
very large. The issue concerns all kinds of rotors 
that are bound by any type of forces that resemble 
a string in their effect of binding.

Let us first analyze our first, simple example.
According to Eq.~(\ref{muconst-p0}), as we increase
$\Omega$ and keep all other quantities constant,
the modulus of the tension in the center, $|p_0|$,
also increases. It continues so till at some point $p_0$ reaches
$-\rho c^2$. At that point, $p_\sigma = \text{const} =
p_0$ and $\epsilon+p=0$ in every point on the string.
If we further increase $\Omega$, then $\epsilon+p$
becomes negative.

A negative $\epsilon+p$ implies that in some frame
of reference the density of energy of the string,
given by the $T^{00}$ component of the energy-momentum 
tensor, is negative, which appears unphysical. Therefore,
it seems that the string ought to break before reaching
such large $\Omega$. Hooke's string also admits
$\epsilon+p<0$, which is the case when the constant 
$C$ is negative.

There exist no ordinary strings that could withstand
elongations big enough to get close to the unphysical
region of $\epsilon+p<0$ prior to the breaking. However,
there may exist microscopic systems that are close to
the case $\epsilon+p=0$. Such systems are expected to 
be found in mesons, built as pairs of quarks connected 
by the gluon string. In such models of mesons, quarks 
are light and move with the speed close to the speed 
of light. The string satisfies the condition $\epsilon+p 
= 0$. This condition is adopted in the mathematical 
string theory, where the massive endpoints are not 
included.

\subsection{Remark on Lagrangian approach} %

Interested readers may ask how the string in our
rotor is related to the strings in theories one
can find in literature. The latter are typically
given in terms of a Lagrangian density, such as
in the case of Nambu-Goto\cite{Nambu,Goto} string or
Chodos-Thorn\cite{ChodosThorn} string. As a matter
of fact, a general motion of a string is such that
each and every infinitesimal bit of the string can
be described as rotating around its center-of-mass
that is momentarily moving with some velocity. So,
our elementary rotor analysis is a necessary ingredient
in building a theory of generally moving strings.
It turns out that the string dynamics one arrives at
using our rotor picture for infinitesimal parts of
the string can be described using a Lagrangian.
Moreover, the Lagrangians one obtains form a class
that includes the Nambu-Goto and Chodos-Thorn
Lagrangians. These results will be presented elsewhere.

\section{Conclusion}
\label{sec:conclusions}

Our relativistic analysis of an elementary example
of an extended physical system, a rotor, allows us
to draw several conclusions. Energy and momentum of
the rotor are not conserved in some inertial frames
of reference unless one properly includes the energy
and momentum that are carried by the string that binds
the endpoint masses. The stresses in the string
contribute to the energy and momentum of the rotor.

The energy contributed by the stresses cancels
the oscillations of energy of a moving rotor obtained 
without proper inclusion of the string. The momentum
induced by the stresses is approximately antiparallel
to the velocity of the rotor and cancels oscillations
in the total momentum of the rotor obtained without
account for the momentum due to the stresses.

We wish to stress that the energy-momentum tensor
contributions that stabilize the total momentum of
the rotor do not manifest themselves in the rest
frame of the system. The stresses that provide the
rotor binding forces become visibly missing in the
frames of reference that move with respect to the
rotor center of mass.

Despite the limitation to a specific system, our
analysis is general enough to say that, a relativistic
description of motion of an extended physical system
with bounded internal motion of its parts, necessitates
inclusion of the dynamics responsible for the binding.
This is true regardless of the binding mechanism --
the total energy and momentum of constituents alone
oscillate in frames of reference that uniformly move
with respect to the system. The energy and momentum
carried by the binding mechanism counter these oscillations.

\appendix
\section{ The four-momentum conservation
                       and its four-vector nature } %

The four dimensional Gauss's theorem applied
to tensor $T$, has the form,
\beq
\int_\Sigma dS_\nu\ T^{\nu\mu}
\es
\int_V d^4 x\  \partial_\nu T^{\nu\mu}
\ ,
\label{gauss-theorem}
\eeq
where $V$ is any four-dimensional region in spacetime. $\Sigma$
is it's boundary, a three-dimensional hypersurface. $dS_\nu$ is
a four-vector perpendicular (in a four-dimensional sense)
to an infinitesimal element of $\Sigma$, pointing outward of
region $V$. Its (four-dimensional) length is a measure
of (three-dimensional) volume of that infinitesimal element.
In particular, if we express $dS_\nu$ in an inertial frame
in which it is perpendicular to a hyperplane of simultaneity,
then $dS_\nu =  n_\nu d^3x$, where $n_0 = \pm 1$. The sign
depends on which direction is the outward one, and $n_i=0$.

\begin{figure}
 \centering
 \includegraphics{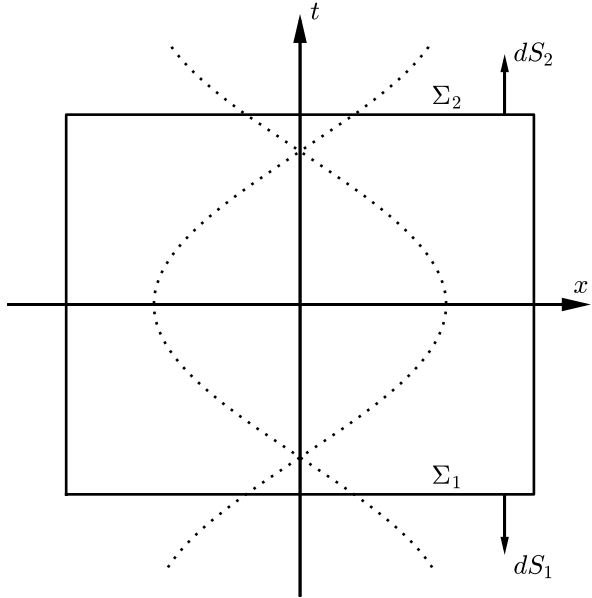}
 \caption{Integration region, which we use to prove
 that divergenceless energy-momentum tensor $T$ gives
 conserved four-momentum $P$. The dotted lines represent
 some spatial extent of a physical system (it does not
 have to be a rotor).}
 \label{fig:const}
\end{figure}

To prove that $P^\mu$ defined in Eq.~(\ref{Pt}) is conserved
in one inertial frame we choose $V$ to be a four-dimensional
cylinder whose axis is the time axis and whose bottom and
top sides, $\Sigma_1$ and $\Sigma_2$ in Fig.~\ref{fig:const},
are located on two hyperplanes of constant time, $t=t_1$ and
$t=t_2>t_1$, respectively. The cylinder is made big enough
in the spatial directions to contain the whole system. Thus,
$T^{\mu\nu}$ is zero on the side surface of the cylinder, as
indicated in Fig.~\ref{fig:const}.
On the bottom hypersurface $\Sigma_1$, the normal outward
unit vector has the time component $n_0=-1$, while on the
top $\Sigma_2$ it has $n_0=1$. The only contributions to
the flux integral on the left hand side of Eq.~(\ref{gauss-theorem})
come from the top and bottom hypersurfaces. These
contributions are equal to $-c P^\mu(t_1)$ and $c P^\mu(t_2)$,
respectively. They sum to zero, assuming that
Eq.~(\ref{conservation-law}) holds. Therefore,
Eq.~(\ref{conservation-law}) implies the four-momentum
conservation.

\begin{figure}
 \centering
 \includegraphics{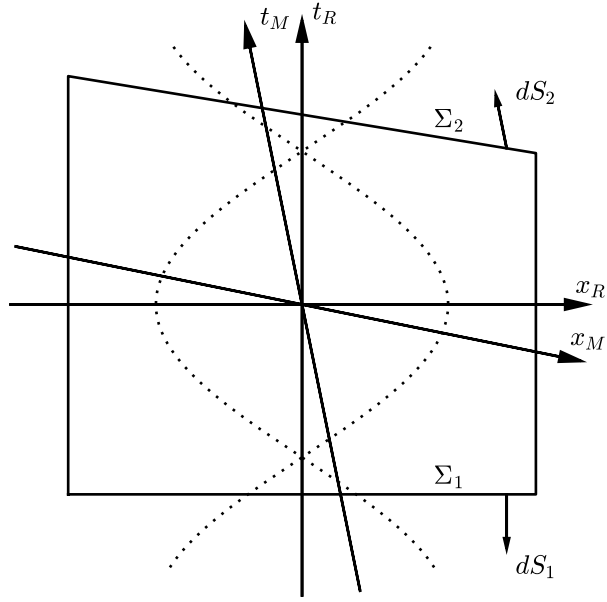}
 \caption{Integration region, which we use to prove that
 four-momentum defined in Eq.~(\ref{Pt}) is a four-vector.}
 \label{fig:pmu4vec}
\end{figure}

Furthermore, if $T$ is a four-tensor then $P^\mu$ is
a four-vector, which means that the extended system's
total four-momenta seen by different inertial observers
are connected by a Lorentz transformation. From the physical
point of view this fact is not trivial because, as we have
seen in Sec.~\ref{sec:2masses}, hyperplanes of simultaneity
are different for different observers. But the Gauss's theorem
is valid no matter how one chooses integration region.
We choose $\Sigma_1$ to lie on the hyperplane of simultaneity
of observer $R$ and $\Sigma_2$ to lie on the hyperplane of
simultaneity of observer $M$. Infinitesimal volume elements
are $dS_1$ on $\Sigma_1$ and $dS_2$ on $\Sigma_2$. Again,
the side boundary of $V$ is located far enough so that it
does not contribute to the left hand side of Eq.~(\ref{gauss-theorem}),
see Fig.~\ref{fig:pmu4vec}.
Now, all steps that are needed to prove that $P^\mu$ is
a four-vector, can be summarized in a sequence of equalities
\beq
P_M^\mu
\es
\frac{1}{c}
\int_{\Sigma_2} d^3 x_M T^{0\mu}_M
\rs
\frac{1}{c}
\int_{\Sigma_2} dS_{2\,M\,\nu} T^{\nu\mu}_M
\rs
-\frac{1}{c}
\int_{\Sigma_1} dS_{1\,M\,\nu} T^{\nu\mu}_M
\nn
\es
-\frac{1}{c}
\int_{\Sigma_1} dS_{1\,M}^\alpha \eta_{\alpha\beta} T^{\beta\mu}_M
\rs
-\frac{1}{c}
\int_{\Sigma_1} ( L^\alpha_{\ \kappa} dS_{1\,R}^\kappa )
\eta_{\alpha\beta}
( L^\beta_{\ \lambda} L^\mu_{\ \nu} T^{\lambda\nu}_R )
\nn
\es
-L^\mu_{\ \nu}
\frac{1}{c}
\int_{\Sigma_1} dS_{1\,R}^\kappa
\eta_{\kappa\lambda} T^{\lambda\nu}_R 
\rs
L^\mu_{\ \nu}
\frac{1}{c}
\int_{\Sigma_1} d^3 x_R T^{0\nu}_R 
\rs
L^\mu_{\ \nu} P_R^\nu
\ .
\eeq
The first equality is a definition of total four-momentum
in the frame $M$, using components of $T$ in that frame.
This is why they carry the subscript $M$. The second equality
makes use of the definition of $dS_{2\,\nu}$. The third equality
is the Gauss's theorem. It equates the total momentum flowing
through hypersurface $\Sigma_2$ to the total momentum flowing
through the surface $\Sigma_1$. Both sides of the Gauss equality are
expressed using coordinates and components in frame $M$.
In the fourth equality, the index of the four-vector $dS_1$
is raised using the Minkowski metric $\eta$.

The fifth equality expresses the components of $dS_1$ and $T$ in
the frame $M$ by their components in the frame $R$ and Lorentz
transformation matrix $L$ from frame $R$ to frame $M$. The sixth
equality makes use of the identity $L^\alpha_{\ \kappa} \eta_{\alpha\beta}
L^\beta_{\ \lambda} = \eta_{\kappa\lambda}$, which is the defining property
of the Lorentz transformation matrices. Written without indices, it
is $L^T \eta L = \eta$.

The seventh equality makes use of the definition of $dS_1$ and the total
four-momentum in the frame $R$. As we see, components $P_M^\mu$ and $P_R^\nu$
are connected via Lorentz transformation and, therefore, define
a four-vector.

Thus, if $T$ is a tensor in special relativity theory, {\it i.e.},
if it transforms from one inertial frame to another by applying
the Lorentz transformation to both its indices, and if its
four-divergence is zero, then total four-momentum defined by
such $T$ in Eq.~(\ref{Pt}) is constant in time and is a Lorentz
four-vector.

\begin{acknowledgments}
	We thank Stanis{\l}aw Ba\.za\'nski, Marek Demia\'nski, 
	Jerzy Kijowski, Andrzej Szymacha, and Jacek Tafel for 
	discussions concerning relativistic description of 
	extended objects.  
\end{acknowledgments}

\end{document}